\documentclass[12pt]{iopart}
\usepackage{graphicx}

\begin{document}

\title[Scaling of the SOL width on MAST]{Scaling of the scrape off layer width during inter-ELM H modes on MAST as measured by infrared thermography}

\author{AJ Thornton, A Kirk and the MAST Team}

\address{EURATOM/CCFE Fusion Association, Culham Science Centre, Abingdon, Oxon, 0X14 3DB, UK}

\ead{andrew.thornton@ccfe.ac.uk}

\begin{abstract}
The power load to the divertor surfaces is a key concern for future devices such as ITER, due to the thermal limits on the material surface. One factor that characterises the heat flux to the divertor is the fall off length in the SOL, which recent empirical scalings have shown could be as small as 1 mm. These predictions are based on a multi-machine scaling of the heat flux width fitted using an expression for the divertor heat flux profile which includes a term for the exponential decay in the scrape off layer (SOL) and diffusion about the last closed flux surface (LCFS) in the private flux region. This expression has been used to fit a database of inter-ELM H mode profiles at the upper divertor and extract the fall off length, $\lambda_q$, for a range of different plasma parameters in double null plasmas. The MAST data shows good agreement with the formula, with the fitted fall off length spanning a range of 5 to 11 mm in the data base generated. Regression of this data has shown that the fall off length has the strongest dependence on the plasma current (or equivalently, the poloidal magnetic field at the outboard midplane) to the power -0.71. The scaling with the smallest $\chi^2$ error utilises the poloidal magnetic field at the outboard midplane ($B_{pol,omp}$) and the power crossing the scrape off layer in the relation $\lambda_q [mm] = 1.84(\pm0.48)B_{pol,omp}^{-0.68(\pm0.14)}P_{SOL}^{0.18(\pm0.07)}$ with $\chi^2=3.46$ and $R^2=0.56$ as a goodness of fit. The equivalent scaling with plasma current is $\lambda_q [mm] = 4.57(\pm0.54)I_{p}^{-0.64(\pm0.15)}P_{SOL}^{0.22(\pm0.08)}$ with $\chi^2=3.84$ and $R^2=0.55$. The moderate goodness of fit suggests that additional plasma parameters are required to accurately reproduce the observed variation in $\lambda_q$.
\end{abstract}

\pacs{52.55.Fa, 52.55.Rk, 52.70.-m} %Check these numbers.

\submitto{Plas. Phys. Control. Fusion}

\maketitle

%Paper starts

\section{Introduction}
\label{section:intro}

The power leaving a tokamak plasma is directed along the scrape off layer (SOL) and onto the divertor surfaces by the magnetic field geometry. The width of the scrape off layer region is of key importance to future devices such as ITER as it sets the size of the parallel heat flux. In ITER it is expected that the power crossing the scrape off layer will be of the order 100 MW \cite{loarte2007}. Initial predictions for ITER based on modelling results have suggested that the inter-ELM H mode fall off length in the scrape off layer, $\lambda _{q}$, will be of the order 5 mm \cite{kukushkin2003}. Recent work \cite{eich2011} has shown that this size could be as small as 1 mm. Taking into account the magnetic geometry of ITER, a heat flux width of 5 mm will give a parallel heat flux of 1 GW m$^{-2}$, which is higher than on present day devices. The divertor is designed to minimise the heat flux to the plasma facing components, via tile inclination and radiation of the power passing along the divertor leg, to levels of the order 10 MWm$^{-2}$ in steady state. In the case where the fall off length is 1 mm, then the heat flux rises significantly and could lead to damage to the divertor materials, limiting the lifetime of the ITER divertor.

The measurement of the fall off length is typically made using infra-red (IR) thermography or Langmuir probe measurements. The measured heat flux is fitted using a range of different techniques to extract the fall off length. Recent work \cite{eich2011, makowski2012, eich2013} has developed a model of the heat flux which corresponds to an exponential decay in the SOL region and the diffusion of power across the last closed flux surface and into the private flux region (PFR) along the divertor leg \cite{wagner1985}. It is this work, using the new model of the heat flux profile, which has suggested that $\lambda _q$ in ITER is of the order 1 mm. In these studies and those that have used only an exponential decay in the SOL region for the fitting \cite{detemmerman2010}, the factor with the strongest effect on $\lambda _q$ is the magnitude of the poloidal magnetic field, $B_p$. It is typically found that $\lambda _q$ scales as $B_p^{\alpha}$ where $\alpha$ ranges from -1.5 to -0.9. 

This paper will use IR data from MAST \cite{meyer2013} to generate a scaling for the inter-ELM H mode fall off length in double null plasmas as a function of various plasma parameters. The fall off length will be determined using the method set out by Eich et al \cite{eich2011} in which the measured heat flux profile, $q$, measured at the upper divertor is fitted with equation \ref{eqn:eich}. The work presented here follows on from work previously performed on MAST using Langmuir probe (LP) data and general agreement is found between the two data sets.

\begin{equation}
	\begin{centering}
	q(\bar{s})=\frac{q_0}{2}exp\left (  \left ( \frac{S}{2\lambda _q \cdot f_x} \right )^2 - \frac{\bar{s}}{\lambda _q \cdot f_x} \right ) erfc \left ( \frac{S}{2\lambda _q \cdot f_x} - \frac{\bar{s}}{S} \right ) + q_{bg}
	\end{centering}
	\label{eqn:eich}
\end{equation}

In equation \ref{eqn:eich}, $\bar{s}=(R-R_{0})$ with $R_{0}$ the location of the last closed flux surface (LCFS), $f_x$ is the flux expansion from the outboard midplane to the divertor surface, $q_0$ is the heat flux at the outboard midplane, $S$ is the power spreading parameter and is the width of the Gaussian and $q_{bg}$ is the background heat flux. All of these parameters are fitted, with the exception of the flux expansion to obtain a value for the heat flux fall off length at the midplane, $\lambda_q$. The flux expansion is determined using the variation of the magnetic field at the divertor and the outboard midplane, as determined by EFIT equilibrium reconstruction \cite{lao1985}.

The structure of the paper will be as follows; section \ref{section:database} will outline the method used to measure the heat flux profiles and the stages used to generate a database of plasma parameters which will be used to determine the scaling. Section \ref{section:scalings} will discuss the scaling of the fall off length with individual upstream plasma parameters and identify the parameters which have the largest effect on setting $\lambda_q$ and act as a guide as to the regressions performed. The means of assessing the quality of the fit from the regression will be described in section \ref{section:quality_reg}. Section \ref{section:regression} will focus on regressing the variables chosen in section \ref{section:scalings} to give a relationship for the fall off length. The error determined from the profile fits will be discussed and verified in section \ref{section:error}. Finally, the paper will conclude with a discussion in section \ref{section:discussion}.

\section{Scrape off layer width database}
\label{section:database}

The database generated for this study focuses on inter-ELM measurements of the upper outer divertor heat flux of attached double null plasmas. The choice of discharge and divertor surface on which to measure the heat flux has been dictated by the availability of IR data during the MAST campaign. The measurements of the heat flux have been taken during a period where only one of the IR cameras was operational, and the use of double null, upper divertor data provides the largest dataset for analysis. In double null plasmas it is possible to generate both ELMing and ELM free discharges, both types of which are included in the database. It should be noted that the strike point footprint broadens during ELMs \cite{eich2011_2} which will affect the calculated value of $\lambda_q$. In order to avoid the effects of ELM broadening, the profiles have been selected which are at least 1.2 ms after an ELM, as measured using divertor D$_{\alpha}$ emission. The duration of 1.2 ms is taken as this is twice the ELM heat flux decay time, as measured in ELMy double null MAST plasmas \cite{thornton2013} and ensures that the plasma has recovered from the ELM.

\begin{figure}[htp]
	\centering
    \includegraphics[width=0.5\textwidth]{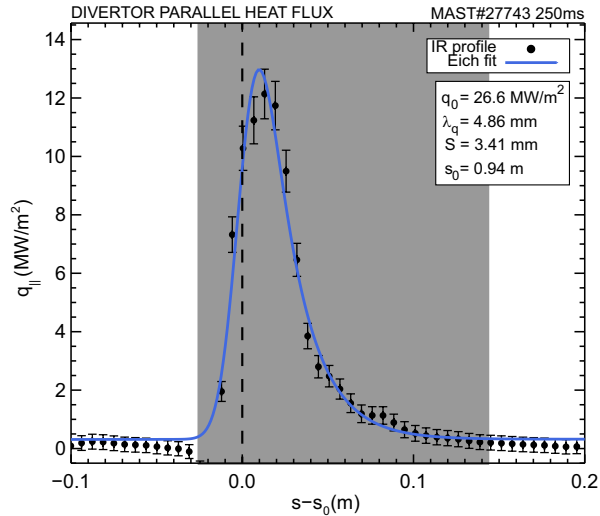}
    \caption{The measured heat flux profile at the upper outboard divertor (black dots) with the fit using the Eich formula (blue curve). The shaded region signifies the region over which the IR profile is fitted to determine the fall off length; a range $3 \cdot S$ wide on the private flux region and $7(S+\lambda_q)$ on the SOL side. The fitted fall off length ($\lambda_q$), parallel heat flux at the midplane, $q_0$, and Gaussian spreading factor, $S$, are shown for this profile.}
    \label{fig:ir_profile}
\end{figure}

The heat flux to the divertor in MAST is routinely measured on MAST using IR thermography \cite{detemmerman2010}. A typical IR profile for the divertor heat flux is shown in \fref{fig:ir_profile} with the measured heat flux converted into the parallel heat flux. The profiles at the selected period are fitted using equation \ref{eqn:eich}, first over the whole profile to enable a suitable range for the fitting to be defined. The initial fit is used to determine the width of the Gaussian and an estimate for the fall off length. The profile is then refitted using a range which is three Gaussian widths wide in the private flux region (PFR) and $7(S+\lambda_q)$ on the SOL side of the profile. These ranges were chosen as it would be expected that three Gaussian widths would encompass 99\% of the profile on the PFR side and the SOL side width was chosen as it provides an acceptable level for the background heat flux ($q_{bg}$).

The fits are filtered based on the chi-square ($\chi^2$) of the fit and then manually inspected to ensure the profiles are accurately fitted. The fitted parameters, along with plasma parameters such as the line integrated density ($n_{line}$), power crossing the scrape off layer determined from power balance ($P_{SOL}$), plasma current ($I_p$), vacuum toroidal magnetic field at the geometric axis ($B^{vac}_{tor}$) and the poloidal field at the outboard midplane ($B_{pol}^{omp}$), major radius of the magnetic axis ($R_{geo}$), Greenwald fraction ($f_{GW}$), safety factor at 95\% of the poloidal flux ($q_{95}$) are stored for each IR profile fitted. Table \ref{table:parameters} shows the ranges in the plasma parameters for the 139 profiles that form the database.

\begin{table}
	\centering
		\caption{Range of plasma parameters included in the database}
		\begin{tabular}{l c c}
			\hline
			Parameter & Range & Units \\
			\hline
			$B_{pol}$ & 0.11-0.23 & Tesla \\
			$B^{vac}_{tor}$ & 0.36-0.41 & Tesla \\
			$I_p$ & 0.45-0.92 & MA \\
			$n_{line}$ & 1.16-2.58 & x$10^{20}$ m$^{-2}$ \\
			$P_{SOL}$ & 0.84-4.42 & MW \\
			$P_{NBI}$ & 0-3.8 & MW \\
			$P_{rad}$ & 0.1-0.7 & MW \\
			$f_{GW}$ & 0.37-0.85 & - \\
			$R_{geo}$ & 0.897-0.991 & m \\
			$q_{95}$ & 4.7-9.1 & - \\
			\hline
		\end{tabular}
		\label{table:parameters}
\end{table}

\begin{figure}[htp]
	\centering
    \includegraphics[width=0.5\textwidth]{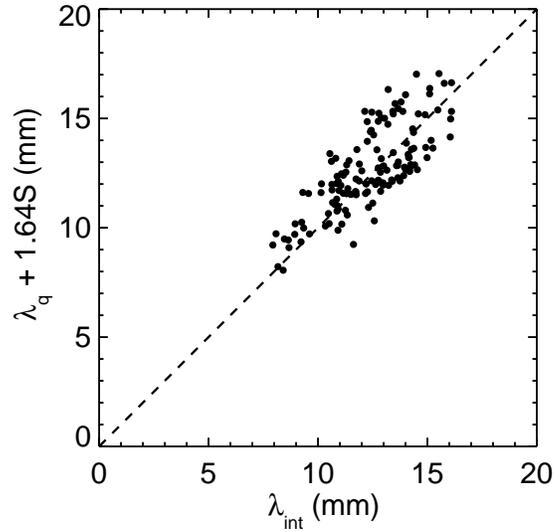}
    \caption{Comparison of the integral heat flux width and the $\lambda_q + 1.64S$ scaling derived from the Eich fit.}
    \label{fig:lambda_int_lambda_q}
\end{figure}

The fall off length from equation \ref{eqn:eich} is one of several methods of quantifying the heat flux width. An alternative method is the integral width, $\lambda_{int}$, and is defined in equation \ref{eqn:int_width} \cite{loarte1999}. It has been shown \cite{makowski2012} that the integral width and the fall off length are related by the formula $\lambda_{int} = \lambda_q + 1.64\cdot S$  when the Eich fit is a good representation of the heat flux profile \cite{eich2013} and that $S/(2\cdot\lambda_q) < 1$ is satisfied \cite{makowski2012}. The relation between $\lambda_{int}$ and $\lambda_q$ provides an additional means of verifying the accuracy of the fit. The integral width is plotted against $\lambda_{int} = \lambda_q + 1.64\cdot S$ in \fref{fig:lambda_int_lambda_q} for the profiles used in the database.

\begin{equation}
	\begin{centering}
		\lambda_{int}= \int_{fit range} \frac{q(s)-q_{bg}}{\mathrm{max}(q(s)-q_{bg})} \mathrm{d}s
	\end{centering}
	\label{eqn:int_width}
\end{equation}

\begin{figure}[htp]
	\centering
    \includegraphics[width=0.5\textwidth]{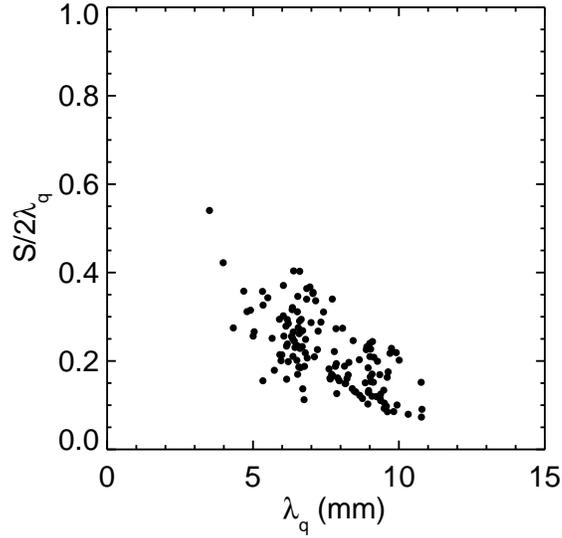}
    \caption{Variation of the Gaussian spreading factor, $S$, with the fitted fall off length $\lambda_q$.}
    \label{fig:lambda_q_vs_s2l}
\end{figure}

\Fref{fig:lambda_int_lambda_q} clearly shows a strong correlation between the fall off length from the Eich formula and the integral width, thereby confirming the quality of the fits used in the database. The requirement for $S/(2\cdot \lambda_q) < 1 $ should also be checked to ensure that the relationship between $\lambda_q$ and $\lambda_{int}$ is applicable to the dataset, which is confirmed by \fref{fig:lambda_q_vs_s2l}.

\subsection{Surface effects on the measured IR profiles}

The measurement of the heat flux to the divertor using IR thermography is affected by the presence of surface layers or surface irregularities. The effects of surface layers on IR measurements have been well documented \cite{herrmann2001, delchambre2009, detemmerman2010} and are seen to generate negative heat fluxes as a result of the rapid heating experienced by a thin layer in good thermal contact with the bulk divertor material. It is important to understand the effects of surface layers on the measured fall off lengths, as this could impact the reliability of the scaling that has been determined. The surface layer coefficient for the divertors on MAST has been determined previously \cite{detemmerman2010} using cross checking between two different wavelength cameras and using energy balance arguments. The optimum value of the surface layer coefficient, $\alpha$ was determined to be 70 kW m$^{-1}$ K$^{-1}$ and this value has been used in this analysis.

\begin{figure}[htp]
	\centering
    \includegraphics[width=0.5\textwidth]{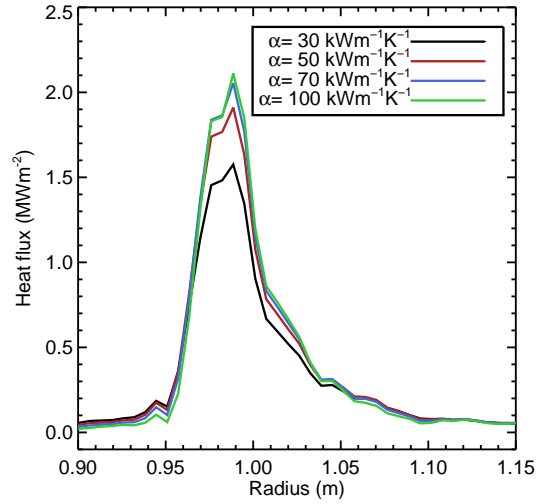}
    \caption{Measured inter-ELM heat flux profiles calculated using a range of surface layer coefficients, $\alpha$.}
    \label{fig:alpha_ir}
\end{figure}

\begin{figure}[htp]
	\centering
    \includegraphics[width=0.5\textwidth]{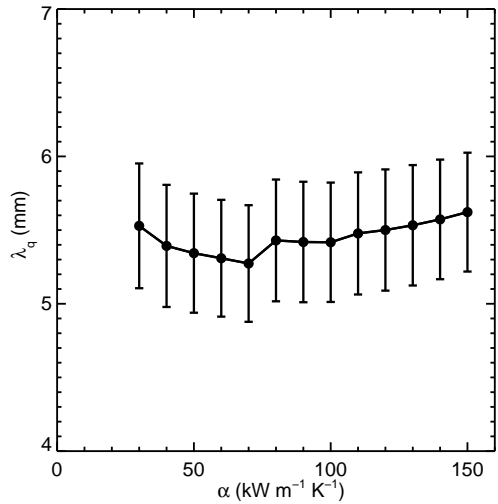}
    \caption{The variation of the fitted fall off length, $\lambda_q$, as a function of the surface layer coefficient $\alpha$.}
    \label{fig:alpha_lambda}
\end{figure}

In order to gain a greater understanding on the effect of the surface layer coefficient of the fall off length, an ELM free discharge was taken and a range of $\alpha$ used from 30 to 200 kWm$^{-1}$K$^{-1}$ to calculate the heat flux to the divertor. \Fref{fig:alpha_ir} shows the heat flux profiles at a given time in the discharge as a function of $\alpha$. It is clear that the surface layer correction has a significant effect on the calculated peak heat flux to the divertor. The profiles in \fref{fig:alpha_ir} are then fitted to determine the fall off length for each of the $\alpha$ values used, which is plotted in \fref{fig:alpha_lambda}. It is clear from \fref{fig:alpha_lambda} that whilst the peak of the heat flux is  affected by the chosen $\alpha$, the fall off length is unaffected with the fitted fall off length for all $\alpha$ lying within the error returned from the fit.

\section{Scaling of the scrape off layer width}
\label{section:scalings}

An effective scaling relationship will only be derived if there is a dependency of the fall off length with the independent parameters which form the basis for the regression. It can be the case that two parameters in the database can be collinear or correlated. In this case, it is necessary to take a selection of the database where one of the parameters is held constant whilst the other varies and then investigate whether or not there is a dependency of the fall off length with the parameter that is allowed to vary. In this section, the correlation between the variables in the database will be investigated and variation with one parameter, holding the others fixed, will be found which will act to guide which parameters can be used in the final regression.

\begin{figure}[htp]
	\centering
    \includegraphics[width=0.5\textwidth]{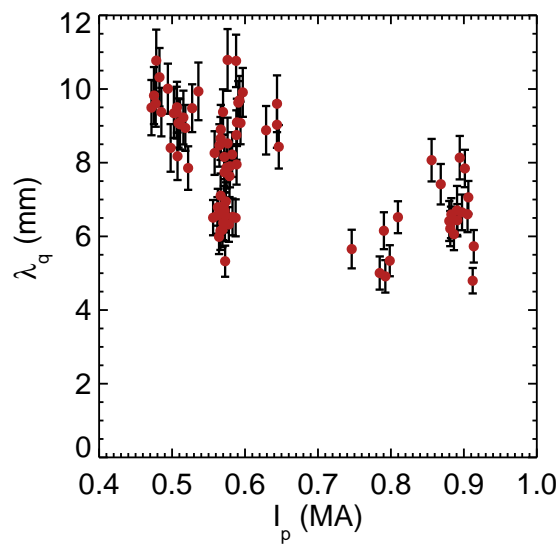}
    \caption{The variation of the fitted fall off length, $\lambda_q$, as a function of the plasma current, $I_p$.}
    \label{fig:ip_vs_lambda_all}
\end{figure}

\begin{figure}[htp]
	\centering
    \includegraphics[width=0.5\textwidth]{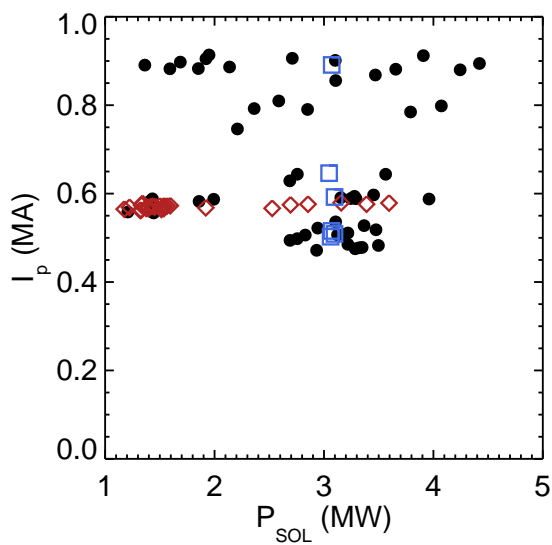}
    \caption{Correlation of the power crossing the scrape off layer ($P_{SOL}$) and the plasma current. The red diamonds correspond to a subset of the data where the $P_{SOL}$ is allowed to vary and the blue squares correspond to a subset of the data where the plasma current is allowed to vary at fixed $P_{SOL}$.}
    \label{fig:psol_vs_ip}
\end{figure}

\begin{figure}[htp]
	\centering
    \includegraphics[width=0.5\textwidth]{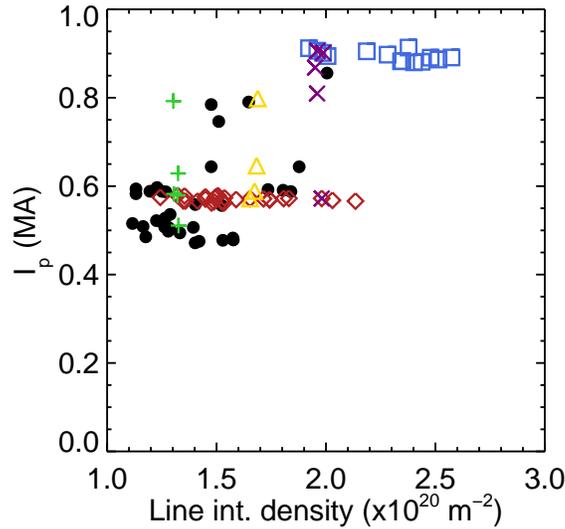}
    \caption{Correlation of the line integrated density in the plasma, as measured by interferometer, and the plasma current ($I_p$). The red diamonds are a subset of this data where the plasma current is constant ($I_p \approx $ 600 kA) and the blue squares are a subset at a higher plasma current of 900 kA. The points shown by the green plus, purple cross and gold triangle are points at fixed density and varying plasma current.}
    \label{fig:density_ip}
\end{figure}

The strongest scaling of the fall off length found in many studies is with the plasma current, $I_p$ \cite{eich2013, gray2011}. The variation of the fall off length with the plasma current is shown in \fref{fig:ip_vs_lambda_all}. As the data shown in \fref{fig:ip_vs_lambda_all} shows all of the data points in the database, there is a range of different plasma densities and power crossing the scrape off layer which gives rise to scatter in the points at a given plasma current. The variation in the density and the power crossing the scrape off layer for the data set is shown in \fref{fig:density_ip} and \fref{fig:psol_vs_ip}. The density of the plasma, as determined from interferometer measurements has been seen to affect the fall off length in other studies on MAST \cite{ahn2006, harrison2013} where Langmuir probe data was used to determine the heat flux to the divertor. These studies have also shown the the power crossing the scrape off layer can have an effect on the measured fall off length. 

The dependence of the fall off length on the plasma current can be investigated be selecting data points at constant $P_{SOL}$ and density and determining the dependency of $\lambda_q$  on plasma current for these points.

\begin{figure}[htp]
    \centering
    \includegraphics[width=0.5\textwidth]{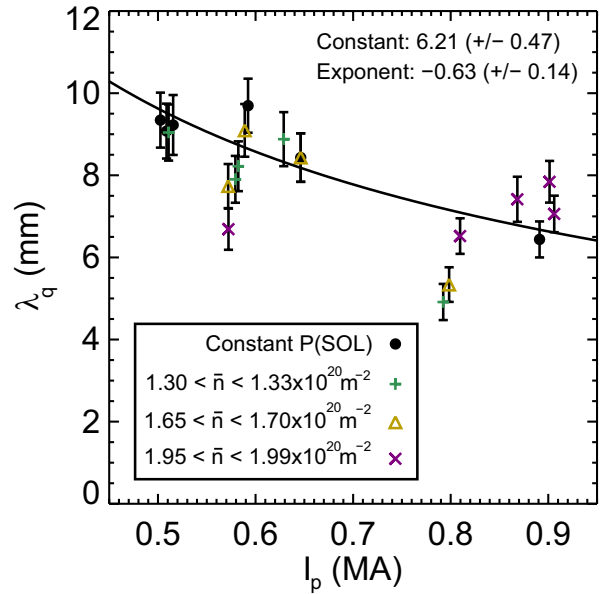}
    \caption{Scaling of the fall off length with the plasma current at a fixed value of $P_{SOL}$ (black circles) and fixed densities (green plus, purple crosses and gold triangles). The fit to the data is of the form $y=A*I_p^{b}$, where A is the constant and b is the exponent shown on the figure.}
    \label{fig:ip_vs_lambda}
\end{figure}

\Fref{fig:ip_vs_lambda} shows the scaling of the fall off length at fixed $P_{SOL}$ (black circles), and for three different selections of fixed density (green plus sign, gold triangle and purple crosses). The constant $P_{SOL}$ values correspond to the blue squares on \fref{fig:psol_vs_ip} and the fixed density data correspond to the points in \fref{fig:density_ip}. The data points here show a clear trend of decreasing $\lambda_q$ for increasing plasma current. Due to the limitations of the data set, it is not possible to chose data points of fixed density and $P_{SOL}$ and as a result scatter from this variation is present in the data in \fref{fig:ip_vs_lambda}. The points at the three different densities support the variation seen at fixed $P_{SOL}$, but they do not show a clear dependence of $\lambda_q$ on the density.

\begin{figure}[htp]
	\centering
    \includegraphics[width=0.5\textwidth]{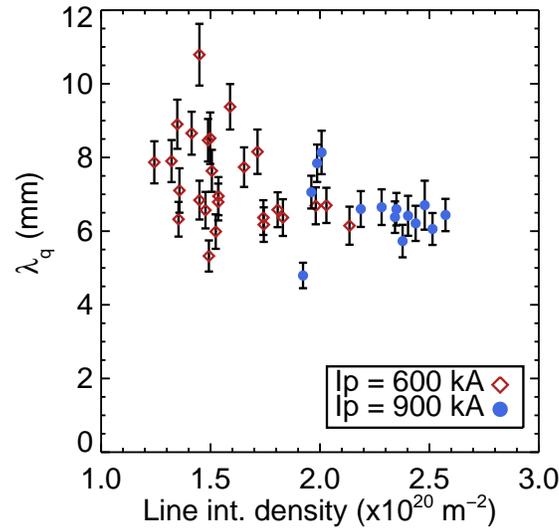}
    \caption{The variation of the fall off lengths, $\lambda_q$, as a function of the line integrated density for two different plasma currents. Data with a plasma current at 600 kA is shown by the red diamonds and data from plasmas with a current of 900 kA is shown by the blue squares.}
    \label{fig:density_lambda}
\end{figure}

The variation of the fall off length with the density can be investigated further by selecting a range of data points at constant plasma current. The chosen points are shown in \fref{fig:density_ip} for a plasma current of 600 kA (red diamonds) and 900 kA (blue squares) respectively. It is clear to see that there is a positive relationship between the plasma current and density which is an artifact of the operation of MAST whereby higher density plasmas are operated at higher current. \Fref{fig:density_lambda} shows the variation of the fall off length with the density for each of these two plasma currents. The dependence of $\lambda_q$ is shown if the average of the 600 kA and 900 kA points are taken, whereby the 600 kA points have a larger $\lambda_q$ than the points at 900 kA. The scaling of the fall off length with the density is not clear from \fref{fig:density_lambda}, there is a weak dependence based on the smaller variation seen in $\lambda_q$ for the higher density points compared to the lower density points. There is a range of $P_{SOL}$ values for the data shown in \fref{fig:density_lambda} which could affect the variation seen. 

%In order to determine if theres is a weak dependence of the density on $\lambda_q$, the $P_{SOL}$ dependence can be removed by dividing the points in \fref{fig:density_lambda} by $P_{SOL}$. The variation of $\lambda_q/P_{SOL}$ is shown in \fref{fig:density_lambda_psol} and shows a trend of decreasing fall off length with density, though the dependence is weak.

%\begin{figure}[htp]
%	\centering
%    \includegraphics[width=0.5\textwidth]{density_vs_lambda_norm_psol}
%    \caption{The variation of the fall off lengths divided by $P_{SOL}$ as a function of the line integrated density for two different plasma currents. Data with a plasma current at 600 kA is shown by the red diamonds and data from plasmas with a current of 900 kA is shown by the blue squares.}
%    \label{fig:density_lambda_psol}
%\end{figure}

The power crossing the scrape off layer is a variable which has been seen to affect the fall off length in several recent studies \cite{eich2013, harrison2013}. In \fref{fig:psol_vs_ip} the power crossing the scrape off layer, $P_{SOL}$, is plotted as a function of the plasma current to determine if there is a correlation between these two parameters. The $P_{SOL}$ value is largely uncorrelated with the plasma current, which permits all of the data collected to be used in the scaling as there is a range of input power for a given plasma current.

\begin{figure}[htp]
	\centering
    \includegraphics[width=0.5\textwidth]{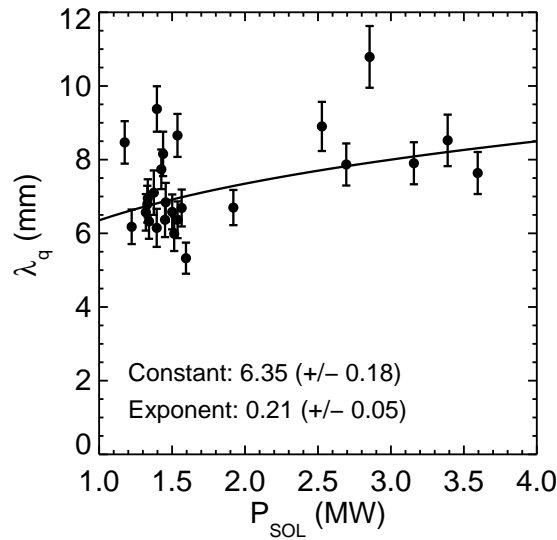}
    \caption{Variation of the fall off length, $\lambda_q$, as a function of the power crossing the scrape off layer, $P_{SOL}$ at a fixed plasma current ($I_p$ = 600 kA). The fit to the data is of the form $y=A*P_{SOL}^{b}$, where A is the constant and b is the exponent shown on the figure.}
    \label{fig:psol_vs_lambda}
\end{figure}

To identify if there is a scaling of the fall off length with $P_{SOL}$, a region of data is selected at constant plasma current as $\lambda_q$ is known to vary with this quantity. The scaling of the fall off length with $P_{SOL}$ can be found using this subset of data (red diamonds in \fref{fig:psol_vs_ip}) and the power crossing the scrape off layer is allowed to vary. The scaling of $P_{SOL}$ with the fall off length is shown in \fref{fig:psol_vs_lambda} and can be fitted using an equation of the form $y=A*P_{SOL}^{b}$. It can be seen from both the data and the fit in \fref{fig:psol_vs_lambda} that there is a weak positive scaling of the fall off length with the power crossing the scrape off layer, which is consistent with previous MAST studies \cite{harrison2013} both in sign and magnitude of exponent and with a recent multi-machine scaling \cite{eich2013}.

For the toroidal magnetic field, $B_t$, one can use either the magnetic field returned from equilibrium reconstruction at the magnetic axis of the plasma, or the vacuum magnetic field at the geometric centre of the tokamak. In the case of the magnetic field returned from equilibrium reconstruction, there is a strong correlation between the toroidal field and the poloidal field at the outboard midplane. The strong correlation and wide variation in the toroidal field at the axis is due to the effect of $\beta_p$ on the plasma. The $\beta_p$ affects whether the plasma is diamagnetic ($\beta_p < 1$) or paramagnetic ($\beta_p > 1$) which in turn changes the toroidal field on axis. In order to remove the interdependence, and remain in line with other SOL width scalings \cite{harrison2013} and confinement scalings \cite{mcdonald2007}, the vacuum field at the geometric axis is used as the toroidal magnetic field. The database contains a limited variation in the vacuum $B_t$. There is some variation in the fall off length with the vacuum $B_t$, however, the data is inconclusive and based on a small sample of the data. Essentially all of the points in the scaling are at the same value of $B_t$, taking this into consideration, along with the results of other studies where the toroidal field is seen to have only a weak scaling with the fall off length \cite{harrison2013, eich2013, makowski2012}, the vacuum $B_t$ will not form part of the regression.

\subsection{Quality of regression}
\label{section:quality_reg}

\begin{figure}[htp]
	\centering
    \includegraphics[width=0.5\textwidth]{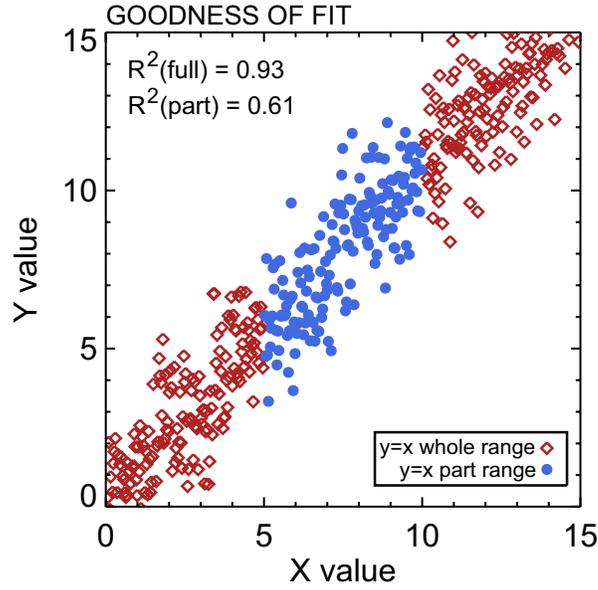}
    \caption{Simulated data of $y=x$ with 7\% noise added to the y values. The $R^2$ goodness of fit is calculated over the full range of the data (red diamonds and blue circles) and a partial range of the data (blue circles) to determine the effect of a reduced data range on the quality of the fit returned by $R^2$.}
    \label{fig:r_sq_testing}
\end{figure}

The quality of the regression is typically assessed using the coefficient of determination ($R^2$), and this quantity is shown for the regressions in table \ref{table:regress}. The $R^2$ coefficient is affected by the number of parameters used in the fit, whereby adding parameters on which $\lambda_q$ is independent results in the $R^2$ value increasing. In addition, there is a large effect on the $R^2$ quality of fit based on the range of values over which the regression is performed. It is an inherent problem with regressions, especially those using data from a single device, that the range of variation is limited and the effect this has on the $R^2$ quality of fit can be illustrated by considering \fref{fig:r_sq_testing}. In \fref{fig:r_sq_testing}, simulated data of the form y=x is used with noise at the 7\% level added to the data to produce some scatter in the points. The 7\% error level is consistent with the error returned from the fitting of the fall off lengths to the profiles. It can be seen from \fref{fig:r_sq_testing} that the $R^2$ value for the full range of x values (red diamonds and blue circles) is 0.93 indicating a good correlation. In contrast, when the same data is fitted, but over a reduced range in the ordinate (blue circles) the quality of fit falls to $R^2=0.61$.

An alternative to $R^2$ to test the goodness of fit is the $\chi^2$ reduced by the degrees of freedom of the fit \cite{squires2001}. The $\chi^2$ value is determined by using the measured $\lambda_q$ as the observed value and the $\lambda_q$ from the regression as the expected value. The error on the fitted $\lambda_q $ is the standard error (one standard deviation) of the fitted parameters. The $\chi^2$ for the regressions are shown in table \ref{table:regress} alongside the $R^2$ values. The $\chi^2$ value and the $R^2$ value are found for each regression and compared in section \ref{section:regression}. 

\section{Regression of SOL width to input parameters}
\label{section:regression}

The regression of the fall off length with the plasma parameters is performed to a power law fit of the form $\lambda_q = C \cdot A^a B^b$ where C is a constant, A-B are the plasma parameters and the powers of the fit are given by a-b. The regression of the parameters is performed in linear space by least squares fitting. Regression in linear space allows better handling of the error on the fitted $\lambda_q$ than performing the regression in log space and has been applied in previous scalings on MAST \cite{harrison2013} and multi-machine datasets \cite{eich2013}. The results from the individual parameter scalings determined in \ref{section:scalings} are used to guide the parameters used in the regression. The fall off length is seen to exhibit a strong scaling with the plasma current, therefore the plasma current is used as a regression variable. Correspondingly, as the plasma current and the poloidal field are related, the plasma current can be replaced in the regression with the poloidal magnetic field at the outboard midplane ($B_{pol}$). The result of the regressions are shown in table \ref{table:regress}, with regressions 1 and 2 showing the difference between regressing against $I_p$ and $B_{pol,omp}$ respectively. A regression against the line integrated density is also included, as a weak dependence of the fall off length on density was seen for the studies of the individual parameters. The error in the regression exponents is returned from the least squares fitting routine used to regress the quantities. The standard error derived from a $\chi^2/dof$ (reduced chi squared) value, where $dof$ is the number of degrees of freedom, assumes that the fit gives $\chi^2/dof=1$. When the $\chi^2/dof$ is large, as is the case for the fitted quantities here, only a small variation in the parameters will cause the reduced chi squared to change by 1, leading to underestimates on the errors in the fitted quantities. In the case of a large reduced chi squared for the fit, the error in the regressed variables can be better estimated for by multiplying the error value returned by the reduced chi square of the fit, it is this value that is quoted as the error in the exponent.

\begin{table}
	\caption{Parameters for each of the regression variables.}
	\footnotesize
	\centering
		\begin{tabular}{l c c c c c c c c}
			\hline
			Reg. & Const. & $B_{pol}^{omp}$ (T) & $I_p$ (MA) & $P_{SOL}$ (MW) & $n_{line}$ x10$^{20}$ m$^{-2}$ & $R^2$ & $\chi^2$ \\
			\hline
			1 & 4.57 ($\pm0.54$) & - & -0.64 ($\pm$0.15) & 0.22 ($\pm$0.08) & - & 0.55 & 3.84 \\
			2 & 1.84 ($\pm0.48$) & -0.68 ($\pm$0.14) & - & 0.18 ($\pm$0.07) & - & 0.56 & 3.46 \\
			3 & 4.03 ($\pm0.88$) & - & -0.75 ($\pm$0.23) & 0.24 ($\pm$0.08) & 0.14 ($\pm$0.19) & 0.56 & 3.81 \\
			\hline
		\end{tabular}
		\label{table:regress}
\end{table}

The regression of the data shows that the strongest dependence is on the poloidal magnetic field at the outboard midplane, however, the difference between the exponent on this regression variable and the plasma current is within the error range on the exponent. The value for the plasma current dependence is within 15\% of the studies previously performed in H mode in MAST when all the data points are included \cite{harrison2013}, but it should be noted that the plasma current range in the data set here is larger than that previously used. There is some variation of the magnitude of the scaling when compared with other machines, especially NSTX \cite{makowski2012, gray2011} which shows a stronger dependence (-1.33 to -1.6 in exponent), though this is over a larger range of plasma current than in the dataset presented in this paper. The $P_{SOL}$ scaling is consistent with previous MAST studies, and those seen across a range of devices \cite{eich2013}. The exponents in the regression match well with the exponents of the fits to the selected data in section \ref{section:scalings}, which confirms that the regression is consistent with the fits when one parameter is varied in isolation. The quality of the fit is moderate for the regressions presented here, the best fit is obtained using the poloidal field and this gives the fall off length to be as shown in equation \ref{eqn:best_fit}. 

\begin{equation}
	\lambda_q [mm] = 1.84(\pm0.48)B_{pol,omp}^{-0.68(\pm0.14)}P_{SOL}^{0.18(\pm0.07)}
	\label{eqn:best_fit}
\end{equation}

Whilst the regression in equation \ref{eqn:best_fit} is the fit with the smallest $\chi^2$ error, the typical scaling parameter for the fall off length is the plasma current. The plasma current is widely used across a range of different machines, and is independent of the location chosen to be the outboard midplane. Therefore, the regression including the plasma current (regression 1, equation \ref{eqn:ip_scal}) should be quoted to allow convenient comparison between the MAST studies presented here and the scalings from elsewhere.

\begin{equation}
	\lambda_q [mm] = 4.57(\pm0.54)I_{p}^{-0.64(\pm0.15)}P_{SOL}^{0.22(\pm0.08)}
	\label{eqn:ip_scal}
\end{equation}

\begin{figure}[htp]
	\centering
    \includegraphics[width=0.5\textwidth]{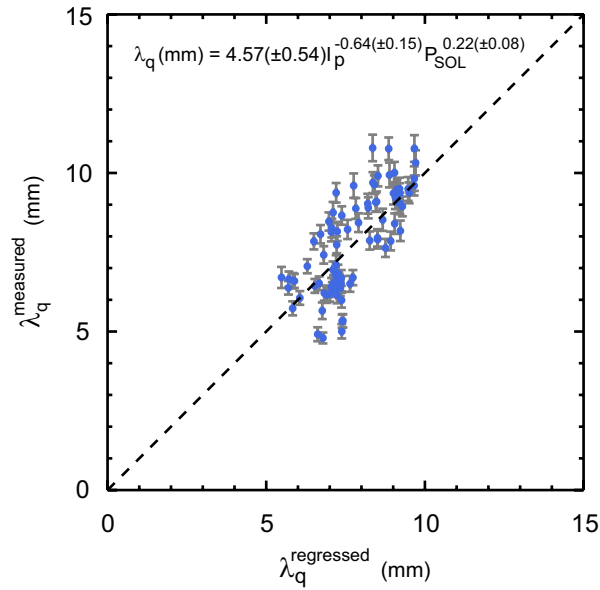}
    \caption{Regression of the measured lambda with most significant plasma parameters ($I_{p}$ and $P_{SOL}$).}
    \label{fig:regressed_lambda}
\end{figure}

The measured fall off length, $\lambda_q^{measured}$ can be plotted against the regressed fall off length, $\lambda_q^{regressed}$, which is shown in \fref{fig:regressed_lambda} to confirm the fitted data accurately represents the measured values.

The regression has also been performed including the density (regression 3). The regression shows a weak dependence with density, with the exponent of the density having the largest relative error of all of the regressed parameters. The fit quality is not improved by the inclusion of the density, remaining similar to the regression including only $I_p$ and $P_{SOL}$. The addition of the density into the regression increases the strength of the scaling with plasma current. This increase is likely due to the operational constraints in MAST by which higher density plasma are performed at higher currents which is consistent with the Greenwald scaling \cite{greenwald2002}. The density dependence from the regression also differs in sign from that expected from the analysis in section \ref{section:scalings}, but is consistent in sign when compared to other studies on MAST \cite{harrison2013}. The magnitude of the density dependence found in these studies is much weaker than seen in past MAST data. These previous studies of the fall off length on MAST \cite{harrison2013, ahn2006} have used Langmuir probes (LPs) to derive the heat flux to the divertor. The derivation of the heat flux from Langmuir probe data requires an assumption to be made about the ion temperature ($T_i$) to determine the sheath heat transmission coefficient. It was assumed in previous studies of the SOL width derived from LP data that the ion ($T_i$) and electron temperatures ($T_e$) were equal, and this is then used to derive the heat flux to the divertor. Measurements of $T_i$ at the divertor have shown that this is not the case \cite{elmore2013}, especially in the case of low collisionality discharges where there is little coupling between the ions and electrons through collisions. The poor coupling at low density leads to ion temperatures higher than electron temperatures. Therefore, assuming that the ion and electron temperatures are the same would cause the profile to be narrower than was actually the case, thereby making the density dependence stronger. Taking into consideration the observation that $T_i/T_e$ decreases with increasing collisionality \cite{elmore2013}, then this would act to broaden the profiles at lower density and lessen the density dependence, thus making the Langmuir probe data more consistent with the SOL width scaling derived from the IR data.

\section{Verification of the fall off error}
\label{section:error}

\begin{figure}[htp]
	\centering
    \includegraphics[width=0.5\textwidth]{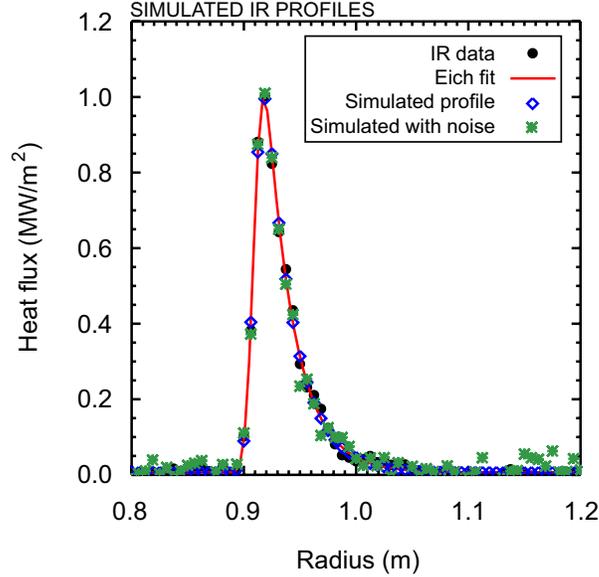}
    \caption{Divertor heat flux profile measured using IR thermography (black dots) with the corresponding Eich fit (red curve). The Eich fit is then used to simulate the IR profile at the same radial locations the IR camera measures (blue diamonds) which gives an exact profile. Noise is then added to the exact profile at a level consistent with the noise on the IR data (7\%) to give a simulated profile (green crosses).}
    \label{fig:simulated_ir}
\end{figure}

The quality of the fits obtained, and the error associated with the $\lambda_q$ derived can be found by fitting simulating profiles with varying amounts of noise. The simulation of the profiles allows the error in the fitted $\lambda_q$ to be verified, which is used in the determination of the quality of the fit (via $\chi^2$). The profile shown in \fref{fig:simulated_ir} (black circles) is measured via IR thermography and fitted using the Eich formula (solid red line). The fitted curve can then be used to simulate experimental profiles with varying levels of noise. The Eich formula is used to generate a simulated IR profile with data points at the radii measured by the IR camera (blue diamonds in \fref{fig:simulated_ir}). The exact simulated profile, for which the correct parameters are known, has varying levels of random noise added to simulate repeated measurement of the profile (\fref{fig:simulated_ir}, green crosses). The level of noise added to the IR profile is the 7\% level seen experimentally, and 1000 simulated profiles are generated to produce sufficient statistics to determine the deviation in the fitted fall off length.

\begin{figure}[htp]
	\centering
    \includegraphics[width=0.5\textwidth]{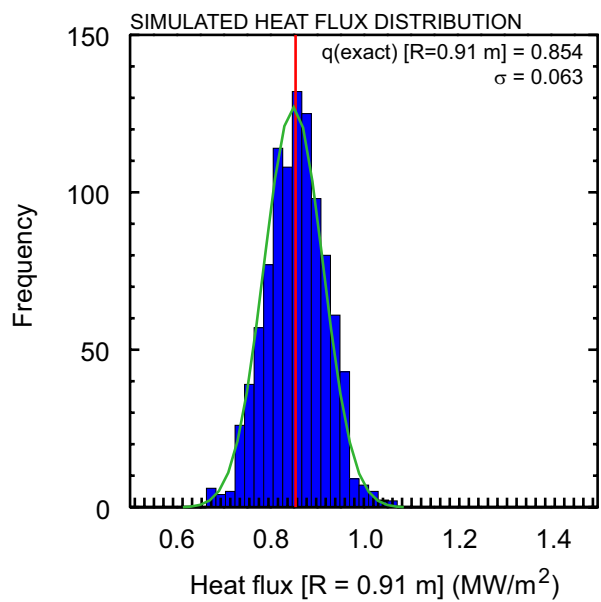}
    \caption{Heat flux extracted from the simulated IR profiles at a radius of 0.91m. The width of the simulated profile is equivalent to the 7\% noise level imposed on the simulated data, which confirms that the errors on the simulated data are Gaussian and evenly distributed about the mean. The exact value (from a simulated curve) is shown by the red vertical line and Gaussian is fitted to determine the width (green curve).}
    \label{fig:check_simulated_errors}
\end{figure}

The variation of the heat flux at a given radial location must be Gaussian for this approach of determining the deviation on the fitted fall length. A histogram showing the variation in the heat flux at a given point in the profile (R=0.91 m) is shown in \fref{fig:check_simulated_errors}. It can be seen that the variation of the heat flux at the radial location is well matched to a Gaussian, with the width of the Gaussian distribution being consistent with the error of 7\% applied to the data. Therefore, the simulated profiles accurately represent making the measurement of the heat flux a repeated number of times and can be used to infer the deviation in fitted values of the fall off length.

\begin{figure}[htp]
	\centering
    \includegraphics[width=0.5\textwidth]{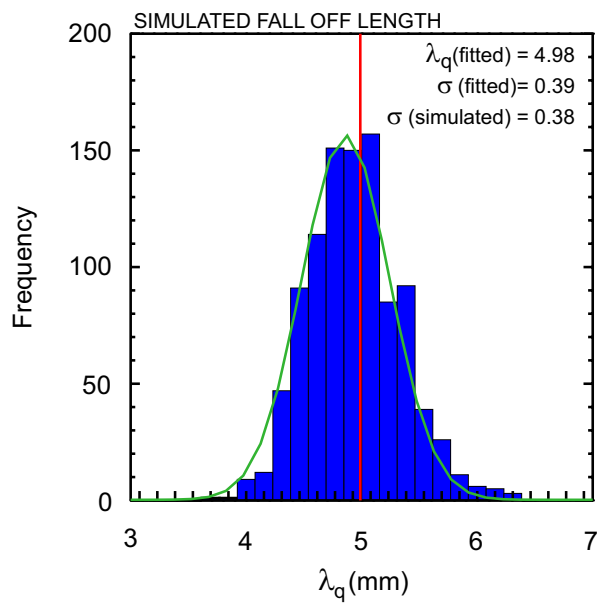}
    \caption{Distribution of the fall off length calculated by fitting the individual profiles. The exact value (from a simulated profile) is shown by the red vertical line and a Gaussian is fitted (green curve) to extract the width. The error returned from fitting the Eich formula to the raw IR data is shown on the plot for comparison with the modelled value.}
    \label{fig:lambda_simulated}
\end{figure}

\Fref{fig:lambda_simulated} shows the fitted fall off length for the 1000 simulated profiles. The exact value of the fall off length input to the simulated profiles is $\lambda_q = 4.98$ mm which is matched by the peak in the histogram. The standard error on the fitted fall off length is $\sigma_{fit} = 0.38$ which is matched well with the standard error from the simulated profiles of $\sigma_{sim} = 0.39$. The good match suggests that the high $\chi^2$ values seen for the regression are due to a parameter missing from the scaling which will require further investigation or improved measurement accuracy to identify.

\section{Discussion}
\label{section:discussion}

The divertor heat flux has been measured on MAST using IR thermography in a range of inter-ELM H mode discharges. The IR measurements of the heat flux profile have then been fitted using the Eich formula, which includes a Gaussian diffusion term to account for diffusion of the heat about the LCFS and an exponential fall off to account for the SOL decay. The Eich formula allows the fall off length at the midplane, $\lambda_q$ to be extracted by accounting for the flux expansion from the midplane to the target. The Eich fall off length has been seen to be related to the integral width, which is widely used elsewhere, by Makowski et al \cite{makowski2012} and the MAST results support this relation which was derived from other conventional tokamaks and NSTX. The confirmation of this result has been used as a means of testing for a good fit to the heat flux profile, and allowing the selection of a reliable dataset for further analysis. An investigation has been performed into the effect of the surface layer coefficient ($\alpha$) on the observed heat flux width, which can have a significant effect on the heat flux profiles returned from IR measurements \cite{detemmerman2010}. The fall off length has been fitted for the same heat flux profiles, using a range of different surface layer coefficients and this has shown that the variation in the fall off length as a result of changing the $\alpha$ is within the error on the fitted fall off length. Hence, the $\alpha$ value is not a significant factor in the determination of the fall off length.

The fall off length has been measured for a range of different plasma parameters such as poloidal magnetic field, toroidal field, density and power crossing the SOL ($P_{SOL}$). The scaling of the fall off length against individual plasma parameters has been used to determine if a given parameter affects the fall off length at all, and to determine the quantities most suitable for regression. The strongest dependence is seen on the plasma current (or equivalently the poloidal magnetic field at the outboard midplane). The fall off length is seen to narrow with increasing plasma current, which is consistent with data from other devices. There is a weak scaling of the fall off length on the power crossing the scrape off layer, in which higher values of $P_{SOL}$ give rise to larger fall off lengths. The dataset does not include sufficient data points at varying toroidal magnetic field to enable an accurate scaling with this quantity to be determined. These dependencies seen in the individual scalings of the fall off length are borne out when a multiple parameter regression is performed across the dataset. The limited range of fall off lengths available when data from one single tokamak is used limits the quality of the fit and this effect has been investigated both using the $R^2$ parameter and $\chi^2$. The analysis has shown that the error on the fit is large and as a result the statistical significance of the fit based on the $\chi^2$ suggest that the probability of a good fit is low. The low confidence level of the good fit suggests an additional parameter is required to fully explain the variation of the fall off length. It is possible that this could be the pedestal electron temperature, however, further work will be required to obtain high resolution Thomson scattering data around the LCFS location to accurately test this hypothesis. Obtaining such measurements of the electron temperature will enable the heat flux fall off width from the IR to be compared to the midplane fall off length in the electron temperature. These quantities should be related from simple SOL physics understanding and from the operating regime, either sheath limited or conduction limited SOL, for a given plasma \cite{stangeby2000}. In addition, it has been seen that inter-ELM filaments arrive at the divertor \cite{ayed2009} and it could be that the arrival of the filaments at the divertor plays a role in determining the $\lambda_q$. Investigation of the arrival of the filaments would require high speed visible imaging of the divertor to determine if a filament has arrived during the integration time of the IR camera and will be the subject of future work.

\ack
This work was part funded by the RCUK Energy Programme [grant number EP/I501045] and the European Communities under the Contract of Association between EURATOM and CCFE. To obtain further information on the data and models underlying this paper please contact PublicationsManager@ccfe.ac.uk. The views and opinions expressed herein do not necessarily reflect those of the European Commission. The authors would also like to thank Dr G. Arnoux and Prof. B Lipschultz for their helpful comments and suggestions.

\section*{References}

\bibliographystyle{unsrt}
\bibliography{AJT_sol_widths_refs}

\begin{thebibliography}{10}

\bibitem{loarte2007}
A~Loarte et~al.
\newblock Progress in the {ITER} {P}hysics {B}asis {C}hapter 4: Power and
  particle control.
\newblock {\em Nucl. Fusion}, 47:S203--63, 2007.

\bibitem{kukushkin2003}
AS~Kukushkin et~al.
\newblock Scaling laws for edge plasma parameters in {ITER} from
  two-dimensional edge modelling.
\newblock {\em Nucl. Fusion}, 43:716, 2003.

\bibitem{eich2011}
T~Eich et~al.
\newblock Inter-{ELM} power decay length for {JET} and {ASDEX} {U}pgrade:
  measurement and comparison with heuristic drift-based model.
\newblock {\em Phys. Rev. Lett.}, 107:215001, 2011.

\bibitem{makowski2012}
MA~Makowski et~al.
\newblock Analysis of a multi-machine database on divertor heat fluxes.
\newblock {\em Phys. Plasmas}, 19:056122, 2012.

\bibitem{eich2013}
T~Eich et~al.
\newblock Scaling of the tokamak near the scape-off layer in {H}-mode power
  width and implications for {ITER}.
\newblock {\em Nucl. Fusion}, 53:093031, 2013.

\bibitem{wagner1985}
F~Wagner.
\newblock A study of the perpendicular particle transport properties in the
  scrape off layer of {ASDEX}.
\newblock {\em Nucl. Fusion}, 25:525--36, 1985.

\bibitem{detemmerman2010}
G~De Temmermann et~al.
\newblock Thermographic studies of heat load asymmetries during {MAST} {L}-mode
  discharges.
\newblock {\em Plasma Phys. Control. Fusion}, 52:095005, 2010.

\bibitem{meyer2013}
H~Meyer et~al.
\newblock Overview of physics results from {MAST} towards {ITER/DEMO} and
  {MAST} {U}pgrade.
\newblock {\em Nucl. Fusion}, 53:104008, 2013.

\bibitem{lao1985}
LL~Lao et~al.
\newblock Reconstruction of current profile parameters and plasma shapes in
  tokamaks.
\newblock {\em Plasma Phys. Control. Fusion}, 25:1611, 1985.

\bibitem{eich2011_2}
T~Eich et~al.
\newblock Type-{I} {ELM} power deposition profile width and temporal shape in
  {JET}.
\newblock {\em J. Nucl. Mater.}, 415:S856--9, 2011.

\bibitem{thornton2013}
AJ~Thornton et~al.
\newblock The effect of resonant magnetic perturbations on the divertor heat
  and particle fluxes in {MAST}.
\newblock Submitted to Nucl. Fusion, 2013.

\bibitem{loarte1999}
A~Loarte et~al.
\newblock Multi-machine scaling of the divertor peak heat flux and width for
  {L}-mode and {H}-mode discharges.
\newblock {\em J. Nucl. Mater.}, 266-269:587--592, 1999.

\bibitem{herrmann2001}
A~Herrmann et~al.
\newblock Limitations for divertor heat flux calculations of fast events in
  tokamaks 27th conf. to plasma physics and controlled fusion (madeira,
  portugal).
\newblock In {\em Proc. EPS 2001}, 2001.

\bibitem{delchambre2009}
E~Delchambre et~al.
\newblock Effect of micrometric hot spots on surface temperature measurement
  and flux calculation in the middle and long infrared.
\newblock {\em Plasma Phys. Control. Fusion}, 51:055012, 2009.

\bibitem{gray2011}
TK~Gray et~al.
\newblock Dependence of divertor heat flux width on heating power, flux
  expansion and plasma current in {NSTX}.
\newblock {\em J. Nucl. Mater.}, 415:S360--4, 2011.

\bibitem{ahn2006}
J-W Ahn et~al.
\newblock L-mode {SOL} width scaling in the {MAST} spherical tokamak.
\newblock {\em Plasma Phys. Control. Fusion}, 48:1077--92, 2006.

\bibitem{harrison2013}
JR~Harrison et~al.
\newblock {L}-mode and inter-{ELM} divertor particle and heat flux width
  scaling on {MAST}.
\newblock {\em J. Nucl. Mater.}, 438:S375--8, 2013.

\bibitem{mcdonald2007}
DC~McDonald et~al.
\newblock Recent progress on the development and analysis of the {ITPA} global
  {H}-mode confinement database.
\newblock {\em Nucl. Fusion}, 47:147, 2007.

\bibitem{squires2001}
GL~Squires.
\newblock {\em Practical {P}hysics}.
\newblock Cambridge University Press, 2001.

\bibitem{greenwald2002}
M~Greenwald et~al.
\newblock Density limits in toroidal plasmas.
\newblock {\em Plasma Phys. Control. Fusion}, 44:R27--80, 2002.

\bibitem{elmore2013}
S~Elmore et~al.
\newblock Scrape-off layer ion temperature measurements at the divertor target
  in mast by retarding field energy analyser.
\newblock {\em J. Nucl. Mater.}, 438:S1212--15, 2013.

\bibitem{stangeby2000}
PC~Stangeby.
\newblock {\em The {P}lasma {B}oundary of {M}agnetic {F}usion {D}evices}.
\newblock Taylor and Francis, 2000.

\bibitem{ayed2009}
N~Ben Ayed et~al.
\newblock Inter-{ELM} filaments and turbulent transport in the {M}ega {A}mp
  {S}pherical {T}okamak.
\newblock {\em Plas. Phys. Control. Fusion}, 51:035016, 2009.

\end{thebibliography}

\end{document}